\documentclass[fleqn,usenatbib,letters]{mnras}

\usepackage[T1]{fontenc}
\usepackage{ae,aecompl}

\usepackage{graphicx}
\usepackage{amsmath}
\usepackage{amssymb}

\title{Discovery of a retrogradely rotating neutron star in the X-ray pulsar GX 301$-$2}

\author[J. M\"onkk\"onen et al.]
{Juhani M\"onkk\"onen,$^{1}$\thanks{E-mail: juhemo@utu.fi}
Victor Doroshenko,$^{2,3}$
Sergey S. Tsygankov,$^{1,3}$\newauthor
Armin Nabizadeh,$^{1}$ 
Pavel Abolmasov$^{1,5}$ 
and Juri Poutanen$^{1,3,4}$
\\
$^1$Department of Physics and Astronomy, FI-20014 University of Turku, Finland\\
$^2$Institut f\"ur Astronomie und Astrophysik, Universit\"at T\"ubingen, Sand 1, D-72076 T\"ubingen, Germany\\
$^3$Space Research Institute of the Russian Academy of Sciences, Profsoyuznaya Str. 84/32, Moscow 117997, Russia\\
$^4$Nordita, KTH Royal Institute of Technology and Stockholm University, Roslagstullsbacken 23, SE-10691 Stockholm, Sweden\\
$^5$Sternberg Astronomical Institute, Moscow State University, 
Universitetsky pr., 13, Moscow 119234, Russia
}

\date{Accepted XXX. Received YYY; in original form ZZZ}

\pubyear{2020}

\begin{document}
\label{firstpage}
\pagerange{\pageref{firstpage}--\pageref{lastpage}}
\maketitle

\begin{abstract}
We report on the analysis of the spin evolution of a slow X-ray pulsar GX~301$-$2 along the orbit using long-term monitoring by {\it Fermi}/GBM. Based on the observationally confirmed accretion scenario and an analytical model for the accretion of angular momentum we demonstrate that in this system, the neutron star spins retrogradely, that is, in a direction opposite to the orbital motion. This first-of-a-kind discovery of such a system proves the principal possibility of retrograde rotation in accreting systems with suitable accretion torque, and might have profound consequences for our understanding of the spin evolution of X-ray pulsars, estimates of their initial spin periods, and the ultimate result of their evolution.
\end{abstract}



\section{Introduction}

GX~301$-$2 is one of the brightest persistent X-ray pulsars, where the neutron star (NS) is accreting from highly structured stellar wind of the hypergiant companion Wray 977 \citep{White1976,Kaper2006}. The donor star has the mass of 39--53 $M_{\odot}$ and the radius $R_*\sim$62 $R_{\odot}$, and loses mass at a high rate ($\sim10^{-5}M_{\odot}$\,yr$^{-1}$) through a slow wind ($\sim300$\,km\,s$^{-1}$) powering the observed X-ray emission from the pulsar \citep{Kaper2006}. The distance to the system 
is estimated to be $d=3.53_{-0.52}^{+0.40}$~kpc based on {\it Gaia} parallax measurement \citep{Treuz2018,BailerJones2018}.
The high mass loss-rate from the primary and the low wind velocity enable
high observed X-ray luminosities reaching $10^{36}-10^{37}$\,erg\,s$^{-1}$ along the highly eccentric ($e\sim 0.46$) orbit with a period of 41.482 $\pm$ 0.001 d \citep{Koh1997, Doroshenko2010}. The orbital light curve exhibits
strong peak near the periastron at $p\approx6.5\times 10^{12}$\,cm, which has been attributed to the enhanced wind density (gas stream) associated with tidal interaction of the wind and the NS \citep{1988MNRAS.232..199S,Leahy2008}. 

The high accretion rate implies also that significant angular momentum should be constantly accreted to the NS which then can be expected to spin up steadily. 
However, the observed long spin period of the source implies that this is clearly not the case. 
This suggests that either the accretion torque is overestimated, there are episodes of spin-down, or an efficient mechanism of angular momentum loss must exist. Such a mechanism can be more easily realised in the presence of a strong magnetic field and therefore it has been suggested that GX~301$-$2 might host an accreting magnetar \citep{Doroshenko2010}. Alternatively, the angular momentum loss may be attributed to a magnetized wind of a primary \citep{Ikhsanov2012} or the presence of a quasi-statical spherical shell around the NS magnetosphere \citep{Shakura2015}. 
We note that the problem is not limited to GX~301$-$2 because a significant fraction of the X-ray pulsars have long spin periods and exhibit long term spin-down trends \citep{2005A&A...442.1135L,2006A&A...455.1165L}. The accretion torque is also expected to align the spin of the NS with its orbital motion, an assumption which usually remains unquestioned, even though it has not been yet verified directly.
We emphasize that the spin evolution of X-ray pulsars is of interest not only when studying the astrophysics of these systems. For instance, accreting NSs are one of the key progenitors of gravitational wave sources, and the lack of a strong prior for the pre-merger NS spin leads to large uncertainties when interpreting observed waveforms and kilonova signatures \citep{2019PhRvD.100l4042E}. 

The problem of spin evolution in X-ray pulsars is far from being fully understood, especially for wind-accreting systems. It is believed, however, that even in wind-accretors, the angular momentum influx is on average proportional to the accretion rate, which is the reason why the existing investigations have so far focused on modelling of the luminosity dependence of the observed spin-up rate \citep{Doroshenko2010}. 
On the other hand, the observed orbital evolution of flux in GX~301$-$2 and other wind accreting systems is clearly related to inhomogeneous wind structure, which is also expected to affect spin evolution \citep{1997ApJ...478..723B}. Yet, this possibility has been largely ignored in past investigations due to the lack of observational data to test the models. 

Being one of the brightest X-ray pulsars, GX~301$-$2 is routinely monitored by all-sky X-ray monitors which have been measuring both X-ray flux and spin frequency since 2008. This makes GX~301$-$2 an ideal target for an investigation of the spin evolution as a function of the orbital phase with specific interest in the periastron passage where a prominent flare occurs regularly when the source passes through a stream of denser wind. 

In this paper, we begin by describing our analysis of the archival spin frequency measurements for GX~301$-$2 and the result for the spin frequency evolution with the orbital phase in Sect.~\ref{sect:data}. In Sect.~\ref{sect:discussion}, we interpret our results in terms of a simple analytical model for the  angular momentum accreted from an inhomogeneous wind and discuss the implications both for the GX~301$-$2 and other accreting NS systems. 
We conclude in Sect.~\ref{sect:concl}.

\section{Data analysis and results}
\label{sect:data}

\begin{figure}
\begin{center}    
\includegraphics[width=0.95\columnwidth]{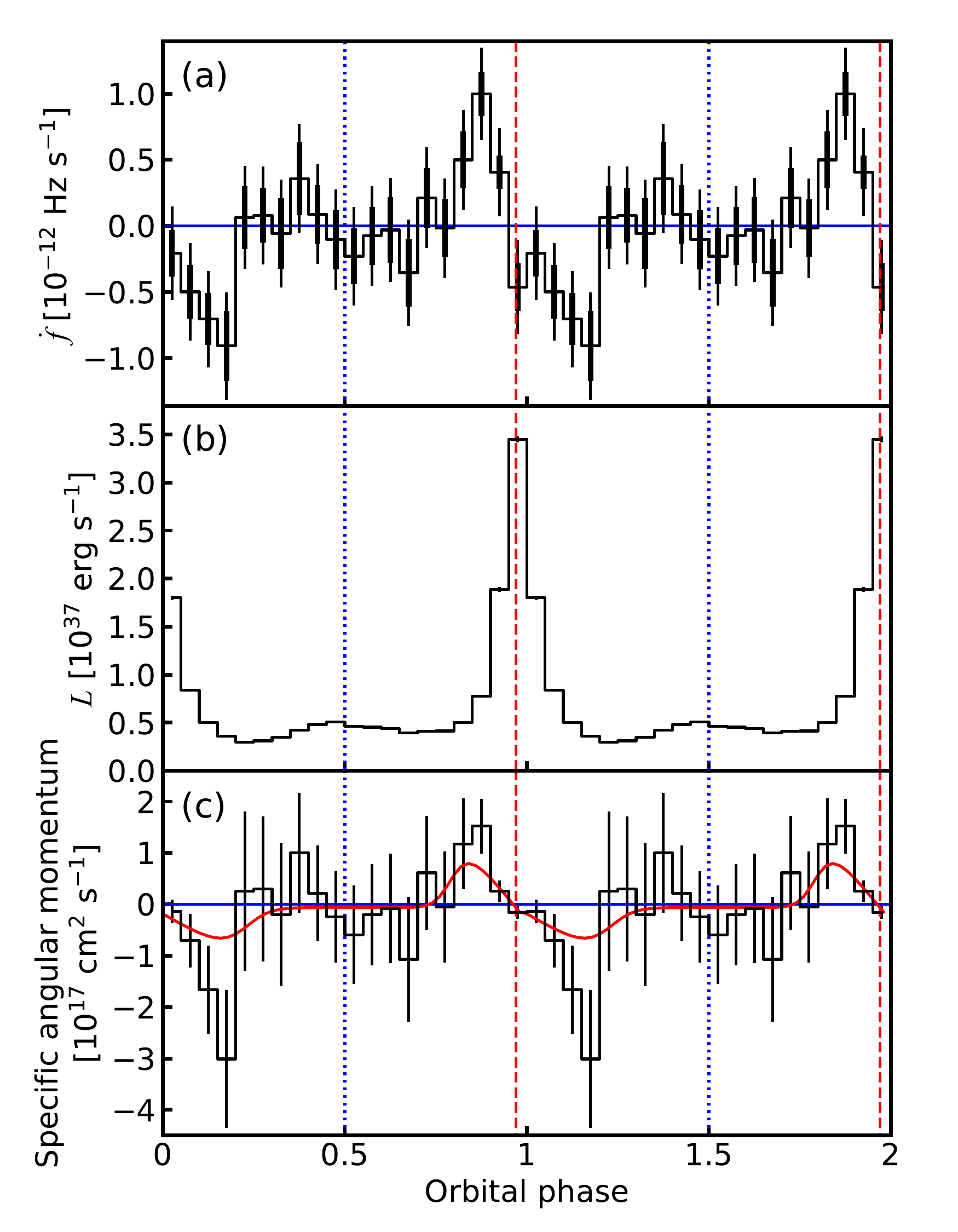}
\end{center}
\caption{(a) Derivative of the pulse frequency of GX~301$-$2 folded on the binary orbital phase. The blue horizontal lines show the zero level while the red dashed and the blue dotted vertical lines in all the panels show the phases of the preperiastron flare and the apastron passage respectively. Thicker error bars indicate the error without the addition of systematics related to orbital uncertainties.
(b) X-ray luminosity folded on the orbital phase. The errors are the one-sigma standard deviations but some are too small to be seen in the plot. (c) Specific angular momentum carried by the accreted matter calculated from the frequency derivative and the estimated mass accretion rate. The red curve shows the model fit to the data described in Sect.~\ref{sect:discussion}.  }
\label{fig:orbital_fold} 
\end{figure}

To investigate the orbital dependence of spin frequency changes we used data from \textit{Fermi} Gamma-Ray Burst Monitor (GBM) Pulsar Project\footnote{\url{http://gammaray.nsstc.nasa.gov/gbm/science/pulsars/lightcurves/gx301m2.html}} from MJD~54693 to 58575. The chosen GBM spin history shows three distinct spin-up episodes \citep{2010ATel.2712....1F,Nabizadeh2019} which were excluded from our data. GBM measures the spin frequency on $\sim2$ day intervals, and the frequency values of two consecutive points were compared to calculate the frequency derivatives. The measurement time of the frequency derivative was then taken as the middle point between these two data points. A total of 1377 data points on 93 orbits were used in the analysis. We verified that the results are not affected significantly if the spin frequency derivative is estimated using other approaches such as numerical differentiation of the interpolated spin history or linear fit to more than two points. 
We then folded the observed spin frequency derivatives $\dot{f}$ with the orbital period of 41.482$\pm$0.001~d \citep{Doroshenko2010}, and calculated the mean and standard deviation of the individual measurements obtained within given orbital phase bin (see Fig.~\ref{fig:orbital_fold}a).

A possible explanation for the observed spin frequency derivative variations would be systematic errors associated with the uncertainty in the orbital parameters of GX~301$-$2. To assess their possible effects on the measured frequency derivative, we first reverted the binary correction applied to GBM data, and then corrected the resulting ``raw'' spin history using  $10^4$ sets of randomly sampled orbital parameters generated around initial values assuming uncertainties reported by \cite{Koh1997}.
We then repeated the analysis described above for each realisation and calculated for each phase the standard deviation of the obtained spin frequency derivatives. This turned out to be $0.5-3\times10^{-13}$\,Hz\,s$^{-1}$ or $\sim0.3-3$ of the statistical uncertainty with the largest values around the periastron, and the maximal value of $3\times10^{-13}$\,Hz\,s$^{-1}$ was added in quadrature to the statistical uncertainty for all calculations below. The original errors and the systematics are shown in Fig.~\ref{fig:orbital_fold}a.
 
To assure that the observed fluctuations of frequency derivative around the periastron are significant, we used the bootstrap method. In particular, we repeated the folding procedure for $10^4$ randomized spin frequency derivative histories obtained by shuffling the observed one, and for each calculated the $\chi^2$ deviation of the folded frequency derivative from a constant value of zero. We found that the distribution of these calculated $\chi^2$ values is, as expected, also consistent with a $\chi^2$ distribution, described by 14 degrees of freedom, a shift of 1.0 and a scaling of 1.4. From such a distribution, we have a chance probability of $6\times10^{-5}$ of obtaining a deviation equal to or greater than $\chi^2 =62$ as measured for the frequency derivative fluctuation in GX~301$-$2. This implies that the observed fluctuations are highly significant.

The source flux was estimated based on \textit{Neil Gehrels Swift} Burst Alert Telescope (BAT) data\footnote{\url{https://swift.gsfc.nasa.gov/results/transients/GX301-2/}} and the X-ray light curve was folded similarly to the frequency derivative to trace the corresponding flux. The flux was transformed to X-ray luminosity using three simultaneous observations by the Nuclear Spectroscopic Telescope Array (\textit{NuSTAR}). In the full energy band, 3$-$79~keV, \textit{NuSTAR} observed absorbed luminosities of $(0.24\pm 0.07)\times 10^{37}$, $(0.4\pm 0.1)\times 10^{37}$ and $(1.2\pm 0.3)\times 10^{37}$~erg~s$^{-1}$ on MJD~56959, 57299 and 58545 \citep{Nabizadeh2019}. The luminosities corrected for absorption in the interstellar medium and in the system are consistently about 20 per cent higher. The corresponding count rates measured by \textit{Swift}/BAT on the same dates are $0.017\pm0.003$, $0.020\pm0.003$ and $0.06\pm0.03$~cnt~cm$^{-2}$~s$^{-1}$ which imply an average conversion factor of $(18\pm 1)\times 10^{37}$~erg~s$^{-1}$/~(cnt~cm$^{-2}$~s$^{-1}$). 

With the computed luminosity (Fig.~\ref{fig:orbital_fold}b), we  now obtain variations of the specific accreted angular momentum (Fig.~\ref{fig:orbital_fold}c)  with the orbital phase  using equation 
\begin{equation}
    l = \frac{2\pi\dot{f}I}{\dot{M}},
\end{equation}
where $I=10^{45}$~g~cm$^2$ is the moment of inertia of the NS and $\dot{M}$ is the mass accretion rate, which is connected to luminosity $L$ through $\dot{M}=L/(\epsilon c^2),$ where  $\epsilon=0.15$ is the assumed efficiency. 

It is evident that the spin evolution during the preperiaston flare is complex, showing both spin-up and spin-down episodes. If the mass accretion rate was simply correlated with the transported angular momentum, the NS would be expected to strongly spin up throughout the flare. However, the maximum spin-up is observed already before the observed peak of the flux, when maximal accretion torque is naively expected. The observed spin-up rate, however, starts to steadily decrease several days prior to the flux maximum even though the accretion rate continues to increase. In fact, zero spin-up rate is observed at the peak of the flare, and the NS is already spinning down right after the peak. The maximal spin-down rate is then reached several days after the peak of the flare, still at a high accretion rate.

\section{Discussion}
\label{sect:discussion}

Here we would like to focus on the discussion of the spin frequency changes around the periastron. As discussed in the data analysis section, even considering uncertainties in the orbital parameters determined by \citet{Koh1997}, the observed evolution of the frequency derivative over the orbit has high statistical significance and therefore, requires a physical explanation. First of all, it is important to emphasize that in GX~301$-$2, the observed orbital flux dependence is neither consistent with accretion from a transient accretion disc, nor from a smooth homogeneous wind. However, in an eccentric binary system, the wind may be enhanced in the periastron due to tidal interaction of the NS with the primary \citep{1988MNRAS.232..199S}. Based on this result, the flaring has been successfully explained with a gas stream model \citep{Haberl1991,Leahy1991,Leahy2002,Leahy2008}. Additionally, modelling of the X-ray absorption column as a function of orbital phase supports this scenario \citep{Leahy2008}. Finally, the presence of the stream has been confirmed directly through near-IR interferometry \citep{2017ApJ...844...72W}, thus this interpretation appears to be well justified.
In this model, the two peaks in the orbital light curve of the source are attributed to the passage of the NS through the stream: the NS overtakes the stream slightly before periastron and is overtaken by the stream at the apastron \citep{Leahy2008}.

\begin{figure}
\begin{center}    
\includegraphics[width=0.90\columnwidth]{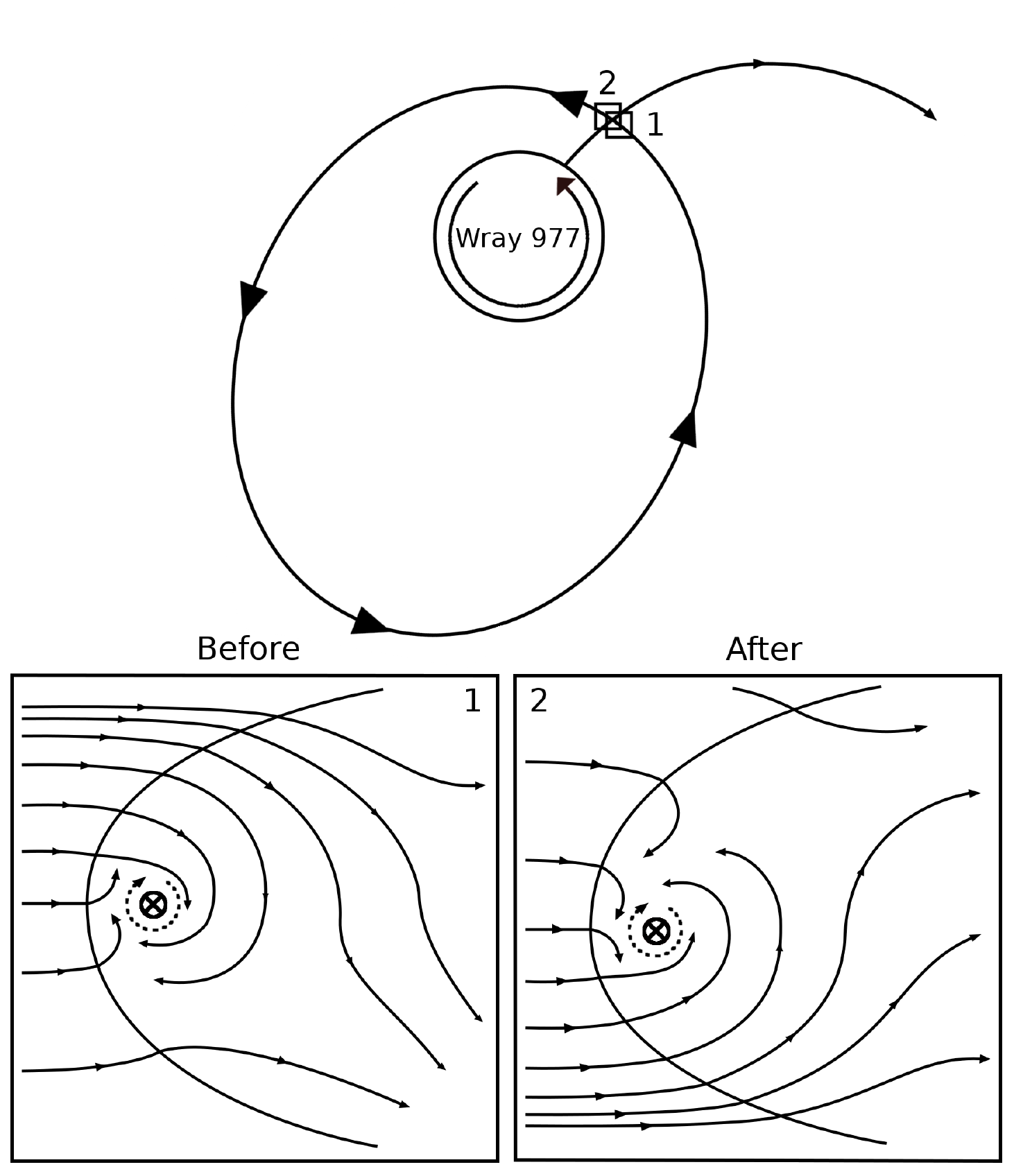}
\end{center} 
\caption{An illustration of the binary system of GX~301$-$2 and the gas stream accretion scenario (not to scale). The NS orbits counterclockwise around the more massive optical companion Wray 977. Right before the periastron, the NS crosses the gas stream, causing the preperiastron flare: insert~1 shows the NS entering the stream and insert~2 shows it exiting the stream. The bow shock is shown and the streamlines illustrate the movement of matter (qualitatively, based on simulation results of \citealt{1999A&A...346..861R} and \citealt{2015ApJ...803...41M}) and the density of the lines indicates the density of matter in the stream. The dotted curve shows the direction of the NS rotation, which is here retrograde with respect to the orbital motion. When the stream is right ahead of the NS, it provides an accelerating torque, and right after the NS has overtaken it, the stream provides a decelerating torque. } 
\label{fig:system_sketch} 
\end{figure}

Now let us discuss the results using a model for the angular momentum transfer if indeed the NS passes through the stream (see Fig.~\ref{fig:system_sketch} for an illustration). 
For simplicity we consider a NS moving with constant velocity $v_{\rm NS}=\Omega r$ perpendicularly to the radial wind of velocity $v_{\rm w}$. 
The wind velocity may depend both on the distance $r$ from the primary star and on the azimuthal angle $\phi$. 
The wind density falls with the radius because of mass conservation $\rho\,v_{\rm w}r^2=$const and it may have a strong azimuthal dependence because of the presence of the gas stream. 
We represent the accreted specific angular momentum as  
\begin{equation} \label{eq:l_acc_total}
l =  \frac{\Omega}{2} R_{\rm A}^2 \eta, 
\end{equation}
where $R_{\rm A}=GM_{\rm NS}/v_{\rm rel}^2$ is the characteristic accretion radius which is a function of the relative velocity, $v_{\rm rel}= \sqrt{v_{\rm w}^2 + v_{\rm NS}^2}$.
The dimensionless coefficient $\eta$ is equal to unity if the wind is fast, homogeneous, spherically symmetric and has a constant velocity \citep{1976ApJ...204..555S}.  
If there are both radial and azimuthal gradients and the wind velocity is not much larger than NS velocity, following \citet{1981A&A...102...36W} we get 
\begin{eqnarray} \label{eq:eta}
\eta &=&   1+ 3\sin^2\alpha + \frac{1+6\cos^2\alpha}{2} \left(\frac{\upartial \ln v_{\rm w}}{\upartial\ln r} \right)  \nonumber \\
&+& \frac{v_{\rm w}}{v_{\rm NS}} 
\left( - \frac{1}{2}\frac{\upartial \ln \rho}{\upartial \phi}  +
3\cos^2\alpha\ \frac{\upartial \ln v_{\rm w}}{\upartial \phi} \right) . 
\end{eqnarray}
where $\alpha$ is the angle between the relative velocity vector and the radial direction, $\sin\alpha= v_{\rm NS}/v_{\rm rel}$. 
This reduces to equation (28b) from \citet{1981A&A...102...36W} if $\alpha\ll 1$, when the azimuthal gradients dominate. 
For spherically symmetric accelerating wind of the form $v_{\rm w}(r)=v_{\infty} (1-R_*/r)^\beta$ 
\begin{equation} 
\frac{\upartial \ln v_{\rm w}}{\upartial \ln r} = 
\frac{\beta} {r/R_* -1 } > 0 
\end{equation} 
and $\eta$ is always positive. 
The radial density gradient produces positive angular momentum, because there is higher density at a trailing part of the accretion cylinder. 
Also the velocity gradient works in the same direction: the wind velocity is smaller behind the NS leading to a larger capture cross-section and a larger arm, resulting also in a positive accreted angular momentum.

The situation is very different if the wind is not homogeneous but has azimuthal gradients of density and velocity.
Let us consider a Gaussian density profile 
\begin{equation} 
\rho(r,\phi)=\rho_0(r) \left[1 + R \exp(-\phi^2/2\sigma^2\right)] 
\end{equation}   
as suggested by \citet{Leahy2008}.  
Here $\phi=0$ is associated with the centre of the stream and the density contrast is defined by the parameter $R$.
In this case we get 
\begin{equation} \label{eq:der_logrho}
\frac{\upartial \ln \rho}{\upartial \phi} = 
- \frac{\phi}{\sigma^2} 
\frac{1}{1+(1/R) \exp(\phi^2/2\sigma^2)} . 
\end{equation}   
This function is positive at $\phi<0$ and negative at $\phi>0$ and has extrema at $\approx\pm2\sigma$ depending on $R$. 
Thus, contribution of the azimuthal gradient of density to  the accreted angular momentum is negative before the NS passes the peak density of the stream (i.e. $\phi<0$), while after the peak, it is positive. 
If the stream velocity is smaller than the background wind velocity then 
the azimuthal derivative $\upartial \ln v_{\rm w}/\upartial \phi$ has an opposite sign compared to the corresponding density derivative, but contributes to the angular momentum with the same sign (see Eq.\,(\ref{eq:eta})). 
Assuming that $\rho v_{\rm w} r^2$ is constant throughout the wind, $\upartial \ln v_{\rm w}/\upartial \phi=-\upartial \ln \rho/\upartial \phi$, and we would get 
\begin{equation} \label{eq:eta_model}
\eta =   1+ 3\sin^2\alpha + 
\frac{1+6\cos^2\alpha}{2} 
\left( \frac{\upartial \ln v_{\rm w}}{\upartial \ln r}  
-  \frac{v_{\rm w}}{v_{\rm NS}}  
 \frac{\upartial \ln \rho}{\upartial \phi} \right)  . 
\end{equation}

Due to a large number of poorly constrained parameters and uncertainty in the velocity structure of the stream, we switch to a more phenomenological approach. We retain the form of  Eq.~\eqref{eq:eta_model}
\begin{equation} \label{eq:eta_phenom}
\eta(\phi) =  C_1 - C_2 \frac{\upartial \ln \rho}{\upartial \phi} (\sigma_1, \sigma_2, R) , 
\end{equation} 
assuming Gaussian azimuthal density profile with the derivative given by Eq.~\eqref{eq:der_logrho} but considering different stream widths on the two sides $\sigma_1$ and $\sigma_2$. 
After all, both the flare and the accreted angular momentum show asymmetry with respect to the flare peak in Fig.~\ref{fig:orbital_fold}.  
The constants $C_1$ and $C_2$ are expected to be of the order of unity, being dependent on the  details of wind properties and the NS orbit.
For GX~301$-$2 the normalization of specific angular momentum in Eq.~(\ref{eq:l_acc_total}) outside the stream is $\Omega R_{\rm A}^2/2 \approx 5\times 10^{16}$~cm$^{2}$~s$^{-1}$ for typical relative velocity of 400\,km\,s$^{-1}$. 
To explain the observed values of $l\sim \pm 10^{17}$~cm$^{2}$~s$^{-1}$, $\eta$ should reach values of about $\pm 2$.

A fit of the model given by Eq.~\eqref{eq:eta_phenom} to the specific angular momentum data is shown in Fig.~\ref{fig:orbital_fold}c. 
Close to the periastron, we can approximate the relation between orbital phase $\Phi$ (mean anomaly) and the azimuthal angle (true anomaly) $\phi$ measured from the stream peak by a linear function $\phi \approx 2\pi s(\Phi-\Phi_{\rm flare})$, where $\Phi_{\rm flare}$ is the phase of the flare and $s$ is a coefficient of the order unity and $s\approx 3$ for eccentricity of $e=0.46$. 
The fluctuation we are interested in occurs near the periastron and therefore we keep the coefficient constant. 
On the other hand, because the stream can itself rotate  around the donor, $s$ should be somewhat smaller. 
Here we do not account for the second crossing of the stream at apastron. 
Fixing the ratio of densities to $R=20$ \citep{Leahy2008} and assuming $s=2$, the best-fit parameters are 
$C_1= -0.1\pm 0.1$, $C_2= -1.0\pm 0.4$,  $\sigma_1= 0.8\pm 0.2$, $\sigma_2= 1.2\pm 0.4$.

At the peak of the flare, associated with the centre of the stream, the accreted specific angular momentum is small (and negative) and the derivative term in Eq.~\eqref{eq:eta_phenom} vanishes. This then determines the smallness of coefficient $C_1$ and its sign in this simplified model. 
There are large values (in absolute sense) of accreted angular momenta in the wings of the flare, well described by the model, which predicts values of the derivative term given by Eq.~\eqref{eq:der_logrho} of the order of $1/\sigma_{1,2}$. 
Interestingly, the observed specific angular momentum is significantly larger after the flare than before. 
For $|C_2|\gg |C_1|$, this behaviour is difficult to explain in terms of this simple model. 
The most important observation is that the signs of the coefficients $C_1$ and $C_2$ are opposite to those expected from theory. 
Indeed, in the wings, the observed values of the accreted angular momentum are of the opposite sign compared to the ones predicted by Eq.\,\eqref{eq:eta_model}. 
This is only possible for a NS which spins retrogradely with respect to the orbital motion.

The best-fit parameters should be taken with caution, as we have ignored the fact that the stream itself rotates around the primary star, resulting in a deviation of the wind velocity from radial direction \citep{1981A&A...102...36W}. 
The ellipticity will also affect the evolution of the relative velocity along the orbit, because the NS radial velocity changes the sign at the periastron. 
This results in a significant decrease of the relative velocity after periastron leading to increase of the accretion radius and the accreted angular momentum. 
This effect might explain the fact that the flare is asymmetric and the absolute value of $l$ is nearly three times larger after the flare than before that. An additional source of error may lie in the uncertainties of the orbital parameters \citep{Koh1997}.

Thus we see that our model describes the angular momentum data well in qualitative terms, although it is difficult to explain the data quantitatively due to the simplifications made and possible uncertainties of the orbit.
In any case, the model demonstrates that the observed spin evolution of the NS can only be explained if it rotates retrogradely with respect to its orbital motion. 
A progradely spinning NS would behave oppositely, i.e. it would spin-down before passing the stream and spin-up after that. 
Although retrograde spins of NSs have been suggested to be common based on the theory of supernova kicks \citep{1983ApJ...267..322H,1995MNRAS.274..461B}, this is the first time it has been observationally demonstrated that such a scenario can indeed be realised. 

We note that the retrograde rotation also naturally explains the long spin period of the source without a need for a strong spin-down torque required to balance the accretion torque. 
While the luminous GX~301$-$2 is the most problematic source in this regard, the problem of long spin periods in X-ray pulsars has in general been a long-standing issue, which retrograde rotation, if found to be more common, can help to address. Retrograde rotation can have profound consequences for the observational appearance of X-ray pulsar population (especially at low mass accretion rates where the expected quiescent luminosity depends on the spin-period of the source; \citealt{2017A&A...608A..17T}), and for the spin distribution of NSs when they merge and produce gravitational waves. So far, there were no observational clues to this problem, and our study represents the first step in this direction. 

Finally, it is worth to note that from a theoretical point of view, the high eccentricity and the low NS-to-donor mass ratio add to the plausibility of a strong supernova kick which resulted in a retrograde NS in GX~301$-$2. The low mass ratio also implies that the primary was likely unable to overfill its Roche lobe and to power a steady disc accretion for extended periods of time, a condition required to align the flipped spin of the NS with its orbital motion. While providing support to our interpretation, these arguments may imply that GX~301$-$2 is a rather special case, and we have to be careful when extending results to other systems.

\section{Conclusions}
\label{sect:concl}

We studied variations of the spin frequency of the persistent X-ray pulsar GX~301$-$2 with the orbital phase.
We found that the spin frequency derivative changes sign around flares when the NS passes through a gas stream close to the periastron. 
These changes are highly significant unless there are major errors in the orbital parameters of the source. 
Using analytical model for the accretion of angular momentum, we showed that the measured changes can only be explained if the NS has a retrograde sense of rotation with respect to its binary orbit. 
The sense of rotation of a pulsar is usually difficult to determine and this first observational evidence that NSs in X-ray binaries might spin retrogradely not only explains the long spin period of the source, but is expected to deepen our understanding of the spin evolution of long-periodic X-ray pulsars in general. Finally, we noted that the possibility of retrograde NSs may have considerable implications for population studies of accreting pulsars as well as studies of supernova kicks and progenitors of gravitational wave sources. 

\section*{Acknowledgements}

This work was supported by the grant 14.W03.31.0021 of the Ministry of Science and Higher Education of the Russian Federation.
We also acknowledge the support from the Vilho, Yrj\"o and Kalle V\"ais\"al\"a Foundation (JM), the Academy of Finland travel grants 317552, 322779, 324550 and 331951 and  the German Academic Exchange Service (DAAD) travel grants 57405000 and 57525212 (VD). 
PA acknowledges the support from the Program of development of Moscow State University (Leading Scientific School `Physics of stars, relativistic objects and galaxies').

\bibliographystyle{mnras} 
\bibliography{library}

\begin{thebibliography}{}
\makeatletter
\relax
\def\mn@urlcharsother{\let\do\@makeother \do\$\do\&\do\#\do\^\do\_\do\%\do\~}
\def\mn@doi{\begingroup\mn@urlcharsother \@ifnextchar [ {\mn@doi@}
  {\mn@doi@[]}}
\def\mn@doi@[#1]#2{\def\@tempa{#1}\ifx\@tempa\@empty \href
  {http://dx.doi.org/#2} {doi:#2}\else \href {http://dx.doi.org/#2} {#1}\fi
  \endgroup}
\def\mn@eprint#1#2{\mn@eprint@#1:#2::\@nil}
\def\mn@eprint@arXiv#1{\href {http://arxiv.org/abs/#1} {{\tt arXiv:#1}}}
\def\mn@eprint@dblp#1{\href {http://dblp.uni-trier.de/rec/bibtex/#1.xml}
  {dblp:#1}}
\def\mn@eprint@#1:#2:#3:#4\@nil{\def\@tempa {#1}\def\@tempb {#2}\def\@tempc
  {#3}\ifx \@tempc \@empty \let \@tempc \@tempb \let \@tempb \@tempa \fi \ifx
  \@tempb \@empty \def\@tempb {arXiv}\fi \@ifundefined
  {mn@eprint@\@tempb}{\@tempb:\@tempc}{\expandafter \expandafter \csname
  mn@eprint@\@tempb\endcsname \expandafter{\@tempc}}}

\bibitem[\protect\citeauthoryear{{Bailer-Jones} et~al.,}{{Bailer-Jones}
  et~al.}{2018}]{BailerJones2018}
{Bailer-Jones} C.~A.~L.,  et~al., 2018, \mn@doi [\aj]
  {10.3847/1538-3881/aacb21}, \href
  {https://ui.adsabs.harvard.edu/abs/2018AJ....156...58B} {156, 58}

\bibitem[\protect\citeauthoryear{{Benensohn}, {Lamb}  \& {Taam}}{{Benensohn}
  et~al.}{1997}]{1997ApJ...478..723B}
{Benensohn} J.~S.,  {Lamb} D.~Q.,   {Taam} R.~E.,  1997, \mn@doi [\apj]
  {10.1086/303835}, \href
  {https://ui.adsabs.harvard.edu/abs/1997ApJ...478..723B} {478, 723}

\bibitem[\protect\citeauthoryear{{Brandt} \& {Podsiadlowski}}{{Brandt} \&
  {Podsiadlowski}}{1995}]{1995MNRAS.274..461B}
{Brandt} N.,  {Podsiadlowski} P.,  1995, \mn@doi [\mnras]
  {10.1093/mnras/274.2.461}, \href
  {https://ui.adsabs.harvard.edu/abs/1995MNRAS.274..461B} {274, 461}

\bibitem[\protect\citeauthoryear{{Doroshenko} et~al.,}{{Doroshenko}
  et~al.}{2010}]{Doroshenko2010}
{Doroshenko} V.,  et~al., 2010, \mn@doi [\aap] {10.1051/0004-6361/200912951},
  \href {https://ui.adsabs.harvard.edu/\#abs/2010A&A...515A..10D} {515, A10}

\bibitem[\protect\citeauthoryear{{East} et~al.,}{{East}
  et~al.}{2019}]{2019PhRvD.100l4042E}
{East} W.~E.,  et~al., 2019, \mn@doi [\prd] {10.1103/PhysRevD.100.124042},
  \href {https://ui.adsabs.harvard.edu/abs/2019PhRvD.100l4042E} {100, 124042}

\bibitem[\protect\citeauthoryear{{Finger} et~al.,}{{Finger}
  et~al.}{2010}]{2010ATel.2712....1F}
{Finger} M.~H.,  et~al., 2010, The Astronomer's Telegram, \href
  {https://ui.adsabs.harvard.edu/abs/2010ATel.2712....1F} {2712, 1}

\bibitem[\protect\citeauthoryear{{Haberl}}{{Haberl}}{1991}]{Haberl1991}
{Haberl} F.,  1991, \mn@doi [\apj] {10.1086/170273}, \href
  {https://ui.adsabs.harvard.edu/\#abs/1991ApJ...376..245H} {376, 245}

\bibitem[\protect\citeauthoryear{{Hills}}{{Hills}}{1983}]{1983ApJ...267..322H}
{Hills} J.~G.,  1983, \mn@doi [\apj] {10.1086/160871}, \href
  {https://ui.adsabs.harvard.edu/abs/1983ApJ...267..322H} {267, 322}

\bibitem[\protect\citeauthoryear{{Ikhsanov} \& {Finger}}{{Ikhsanov} \&
  {Finger}}{2012}]{Ikhsanov2012}
{Ikhsanov} N.~R.,  {Finger} M.~H.,  2012, \mn@doi [\apj]
  {10.1088/0004-637X/753/1/1}, \href
  {https://ui.adsabs.harvard.edu/\#abs/2012ApJ...753....1I} {753, 1}

\bibitem[\protect\citeauthoryear{{Kaper}, {van der Meer}  \& {Najarro}}{{Kaper}
  et~al.}{2006}]{Kaper2006}
{Kaper} L.,  {van der Meer} A.,   {Najarro} F.,  2006, \mn@doi [\aap]
  {10.1051/0004-6361:20065393}, \href
  {https://ui.adsabs.harvard.edu/\#abs/2006A&A...457..595K} {457, 595}

\bibitem[\protect\citeauthoryear{{Koh} et~al.,}{{Koh} et~al.}{1997}]{Koh1997}
{Koh} D.~T.,  et~al., 1997, \mn@doi [\apj] {10.1086/303929}, \href
  {https://ui.adsabs.harvard.edu/\#abs/1997ApJ...479..933K} {479, 933}

\bibitem[\protect\citeauthoryear{{Leahy}}{{Leahy}}{1991}]{Leahy1991}
{Leahy} D.~A.,  1991, \mn@doi [\mnras] {10.1093/mnras/250.2.310}, \href
  {https://ui.adsabs.harvard.edu/abs/1991MNRAS.250..310L} {250, 310}

\bibitem[\protect\citeauthoryear{{Leahy}}{{Leahy}}{2002}]{Leahy2002}
{Leahy} D.~A.,  2002, \mn@doi [\aap] {10.1051/0004-6361:20020781}, \href
  {https://ui.adsabs.harvard.edu/abs/2002A&A...391..219L} {391, 219}

\bibitem[\protect\citeauthoryear{{Leahy} \& {Kostka}}{{Leahy} \&
  {Kostka}}{2008}]{Leahy2008}
{Leahy} D.~A.,  {Kostka} M.,  2008, \mn@doi [\mnras]
  {10.1111/j.1365-2966.2007.12754.x}, \href
  {https://ui.adsabs.harvard.edu/\#abs/2008MNRAS.384..747L} {384, 747}

\bibitem[\protect\citeauthoryear{{Liu}, {van Paradijs}  \& {van den
  Heuvel}}{{Liu} et~al.}{2005}]{2005A&A...442.1135L}
{Liu} Q.~Z.,  {van Paradijs} J.,   {van den Heuvel} E.~P.~J.,  2005, \mn@doi
  [\aap] {10.1051/0004-6361:20053718}, \href
  {https://ui.adsabs.harvard.edu/abs/2005A&A...442.1135L} {442, 1135}

\bibitem[\protect\citeauthoryear{{Liu}, {van Paradijs}  \& {van den
  Heuvel}}{{Liu} et~al.}{2006}]{2006A&A...455.1165L}
{Liu} Q.~Z.,  {van Paradijs} J.,   {van den Heuvel} E.~P.~J.,  2006, \mn@doi
  [\aap] {10.1051/0004-6361:20064987}, \href
  {https://ui.adsabs.harvard.edu/abs/2006A&A...455.1165L} {455, 1165}

\bibitem[\protect\citeauthoryear{{MacLeod} \& {Ramirez-Ruiz}}{{MacLeod} \&
  {Ramirez-Ruiz}}{2015}]{2015ApJ...803...41M}
{MacLeod} M.,  {Ramirez-Ruiz} E.,  2015, \mn@doi [\apj]
  {10.1088/0004-637X/803/1/41}, \href
  {https://ui.adsabs.harvard.edu/abs/2015ApJ...803...41M} {803, 41}

\bibitem[\protect\citeauthoryear{{Nabizadeh} et~al.,}{{Nabizadeh}
  et~al.}{2019}]{Nabizadeh2019}
{Nabizadeh} A.,  et~al., 2019, \mn@doi [\aap] {10.1051/0004-6361/201936045},
  \href {https://ui.adsabs.harvard.edu/abs/2019A&A...629A.101N} {629, A101}

\bibitem[\protect\citeauthoryear{{Ruffert}}{{Ruffert}}{1999}]{1999A&A...346..861R}
{Ruffert} M.,  1999, \aap, \href
  {https://ui.adsabs.harvard.edu/abs/1999A&A...346..861R} {346, 861}

\bibitem[\protect\citeauthoryear{{Shakura} et~al.,}{{Shakura}
  et~al.}{2015}]{Shakura2015}
{Shakura} N.~I.,  et~al., 2015, \mn@doi [Astronomy Reports]
  {10.1134/S1063772915070112}, \href
  {https://ui.adsabs.harvard.edu/\#abs/2015ARep...59..645S} {59, 645}

\bibitem[\protect\citeauthoryear{{Shapiro} \& {Lightman}}{{Shapiro} \&
  {Lightman}}{1976}]{1976ApJ...204..555S}
{Shapiro} S.~L.,  {Lightman} A.~P.,  1976, \mn@doi [\apj] {10.1086/154203},
  \href {https://ui.adsabs.harvard.edu/abs/1976ApJ...204..555S} {204, 555}

\bibitem[\protect\citeauthoryear{{Stevens}}{{Stevens}}{1988}]{1988MNRAS.232..199S}
{Stevens} I.~R.,  1988, \mn@doi [\mnras] {10.1093/mnras/232.1.199}, \href
  {https://ui.adsabs.harvard.edu/abs/1988MNRAS.232..199S} {232, 199}

\bibitem[\protect\citeauthoryear{{Treuz} et~al.,}{{Treuz}
  et~al.}{2018}]{Treuz2018}
{Treuz} S.,  et~al., 2018, arXiv e-prints, \href
  {https://ui.adsabs.harvard.edu/abs/2018arXiv180611397T} {p. arXiv:1806.11397}

\bibitem[\protect\citeauthoryear{{Tsygankov} et~al.,}{{Tsygankov}
  et~al.}{2017}]{2017A&A...608A..17T}
{Tsygankov} S.~S.,  et~al., 2017, \mn@doi [\aap] {10.1051/0004-6361/201630248},
  \href {https://ui.adsabs.harvard.edu/abs/2017A&A...608A..17T} {608, A17}

\bibitem[\protect\citeauthoryear{{Waisberg} et~al.,}{{Waisberg}
  et~al.}{2017}]{2017ApJ...844...72W}
{Waisberg} I.,  et~al., 2017, \mn@doi [\apj] {10.3847/1538-4357/aa79f1}, \href
  {https://ui.adsabs.harvard.edu/abs/2017ApJ...844...72W} {844, 72}

\bibitem[\protect\citeauthoryear{{Wang}}{{Wang}}{1981}]{1981A&A...102...36W}
{Wang} Y.~M.,  1981, \aap, \href
  {https://ui.adsabs.harvard.edu/abs/1981A&A...102...36W} {102, 36}

\bibitem[\protect\citeauthoryear{{White} et~al.,}{{White}
  et~al.}{1976}]{White1976}
{White} N.~E.,  et~al., 1976, \mn@doi [\apjl] {10.1086/182281}, \href
  {http://adsabs.harvard.edu/abs/1976ApJ...209L.119W} {209, L119}

\makeatother
\end{thebibliography}

\bsp	
\label{lastpage}
\end{document}